*Review*

# A Review of Smart Materials in Tactile Actuators for Information Delivery

Xin Xie [1], Sanwei Liu [1], Chenye Yang [2], Zhengyu Yang [3], Tian Liu [1], Juncai Xu [4], Cheng Zhang [5,*] and Xianglin Zhai [6,*]

[1] Department of Mechanical and Industrial Engineering, Northeastern University, Boston, MA 02115, USA; xie.x@husky.neu.edu (X.X.); lsanwei@gmail.com (S.L.); tianliu2010@gmail.com (T.L.)
[2] Department of Electrical Engineering and Computer Science, Massachusetts Institute of Technology, Cambridge, MA 02139, USA; jasony84@mit.edu
[3] Department of Electrical and Computer Engineering, Northeastern University, Boston, MA 02115, USA; yang.zhe@husky.neu.edu
[4] China Institute of Water Resources and Hydropower Research, Beijing Post Code, China; xujc@hhu.edu.cn
[5] Medtronic, Inc., Tempe, AZ 85286, USA
[6] Department of Chemistry, Louisiana State University, Baton Rouge, LA 70803, USA
* Correspondence: cheng.zhang2@medtronic.com (C.Z.); xzhai1@lsu.edu (X.Z.)

**Abstract:** As the largest organ in the human body, the skin provides the important sensory channel for humans to receive external stimulations based on touch. By the information perceived through touch, people can feel and guess the properties of objects, like weight, temperature, textures, and motion, etc. In fact, those properties are nerve stimuli to our brain received by different kinds of receptors in the skin. Mechanical, electrical, and thermal stimuli can stimulate these receptors and cause different information to be conveyed through the nerves. Technologies for actuators to provide mechanical, electrical or thermal stimuli have been developed. These include static or vibrational actuation, electrostatic stimulation, focused ultrasound, and more. Smart materials, such as piezoelectric materials, carbon nanotubes, and shape memory alloys, play important roles in providing actuation for tactile sensation. This paper aims to review the background biological knowledge of human tactile sensing, to give an understanding of how we sense and interact with the world through the sense of touch, as well as the conventional and state-of-the-art technologies of tactile actuators for tactile feedback delivery.

## 1. Mechanoreceptors

Human skin contains a variety of different mechanoreceptors (touch receptors), each with its own structure, placement, frequency response, spatial resolution, adaptation speed, and necessary magnitude of skin indentation to produce a response. Mechanoreceptors are the sensing units that lie underneath the very outer skin surface for retrieving stimulations. The presence and spacing of mechanoreceptors varies between glabrous (naturally hairless) and hairy skin. There are four main types of mechanoreceptors that react to different kinds of stimuli information, such as vibration, shear, texture, and pressure. These four receptors are called Pacinian corpuscles, Meissner's corpuscles, Merkel's discs, and Ruffini endings [1]. Each receptor has a similar basic sensing element, except that their packaging and depth in the skin are adapted to their specific sensing purposes.

Figure 1 shows a 3D diagram of tactile receptors in the skin. The four mechanoreceptors can be divided into two major classifications, fast adapting types and slow adapting types, based on their sensing capabilities. The fast adapting receptors include Meissner's corpuscles and Pacinian corpuscles, and are only sensitive to transitions like vibration and fluttering. The slow adapting receptors include Merkel's discs and Ruffini endings, and they are sensitive to steady or static stimulations, like pure indentation. The receptors can be further identified as type I and type II, based on their depth below the skin's surface. Type I receptors are referred to as Meissner's corpuscles and Merkel's discs; they are shallower receptors located at around 0.7 mm to 0.9 mm below the skin surface. Type II receptors are deeper ones like Pacinian corpuscles and Ruffini endings, which are often located at about 1–2 mm depth below the skin. The deeper receptors have larger receptive area [2].

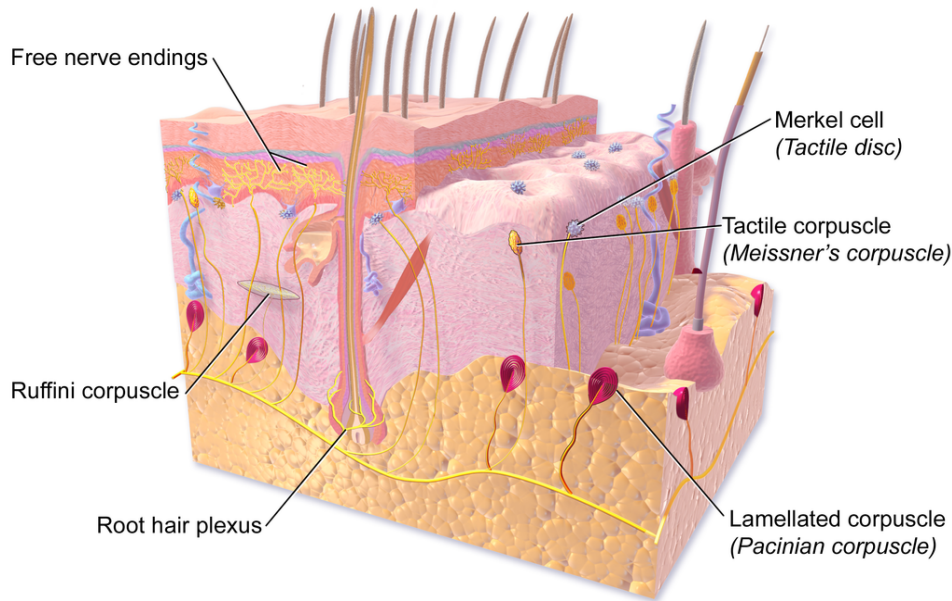

**Figure 1.** Tactile receptors in the skin [3].

**Table 1.** Characteristics of the four mechanoreceptors [4].

| Receptors | Type (Depth) | Type (Adapting Speed) | Receptive Field | Frequency (Hz) | Sensing Property |
|---|---|---|---|---|---|
| Merkel's discs | I | Slow | Small | 5–15 | Pressure, Texture |
| Ruffini endings | II | Slow | Large | 15–400 | Stretch |
| Meissner's corpuscles | I | Fast | Small | 20–50 | Stroke, Fluttering |
| Pacinian corpuscles | II | Fast | Large | 60–400 | Vibration |

Merkel's disks and Ruffini's corpuscles are classified as slowly adapting mechanoreceptors. Pressure, stretch, and static discrimination are sensed by these receptors. Higher force and larger displacement are needed to convey static stimuli sensations [5]. Receptors terminating in Merkel cells are found near the surface of the skin, and have an excellent spatial resolution, with an ability to resolve stimuli separated by as little as 0.5 mm in glabrous skin [6]. Merkel receptors are the primary receptors that are used in reading Braille. However, their best sensitivity to skin indentation is found in the range of 5 Hz to 15 Hz, at which frequency, a minimum skin indentation on the order of 50 μm is typically required to produce a response [7].

Meissner's corpuscles, which are located in the shallowest area below the skin's surface, are particularly suitable to sense low-frequency stroking and vibration. Meissner's corpuscles have a maximum sensitivity between about 20 Hz to 50 Hz, and have a minimum sensitivity to skin indentation of about 14 μm [8]. Meissner's corpuscles are located with a high density of about 150 receptors/cm$^2$, but they have a relatively lower spatial resolution and respond rather uniformly across their entire 3–5 mm receptive field. Pacinian corpuscles differ from Meissner's corpuscles in their shape, depth, and response threshold. They are more sensitive to the vibrations with frequency ranging from a few tens of Hz to hundreds of Hz, with the highest sensitivity around 250 Hz [9]. The highest sensitivity may be found in Pacinian corpuscles, which have demonstrated sensitivity to less than 1 µm skin indentations around 250–300 Hz, and an effective frequency range of about 60–400 Hz [8]. Pacinian corpuscles have a large receptive field, and can sense larger vibrations from a distance of on the order of a centimeter away from the receptor. However, smaller vibrations near the 250 Hz frequency of optimal sensitivity produce a response that is localized directly over the Pacinian corpuscle [9], thereby enabling improved spatial localization with these highly sensitive receptors. The characteristics of mechanoreceptors have a promising degree of overlap with the known characteristics of microactuators and MEMS systems. Pacinian corpuscles also adapt faster than Meissner's corpuscles and have a lower response threshold [10]. For these reasons, Pacinian corpuscles are often considered as the better candidate for retrieving the information primarily about dynamic motion stimuli and high-frequency vibrations.

## 2. Stimulation Methods and Actuation Technologies

Ideally, the display's resolution should leverage the approximately one unit per mm² spacing of mechanoreceptors in human finger pads [11], and be extendable to full 2D. It should be refreshable in real time (i.e., it should refresh at least as quickly as human mechanoreceptors can react), allowing the information conveyed by the display to keep up with rapidly changing inputs. The display should also code information in a way that is easily detected and interpreted, so that it is intuitive and easy to learn to use. The psychophysics literature offers clear insights into humans' ease of sensing various types of tactile stimuli (static vs moving, vibrating vs quasistatic, and low vs high amplitude). In particular, humans are much more sensitive to motions and changing stimuli than they are to static patterns [12], whether those stimuli are visual, audible, or tactile. The display should therefore code information not only as static patterns, but also as simulated motion against the user's finger pads. Finally, its power consumption should be compatible with portable use, and it should be manufacturable by efficiently scalable means, to ensure that its cost is compatible with the resources of its intended user base. There are several ways to provide the stimuli for tactile displays. The stimulation methods can be divided into approximately three categories: thermal, electrical, and mechanical stimuli.

*2.1. Thermal*

Thermal stimulation, or thermal flow, is usually used for adding quality characteristics in the information delivery, such as mimicking the color rendering of a vision system. A 3 × 3 thermal tactile display device, with multiple heat sources that can display temperatures from 5 °C to 55 °C and produce different temperature field distributions, has been developed [13]. Thermal sensation modeling for the fingertip has been developed, showing that skin is more sensitive to rapid temperature change, which leads to a difficulty in presenting long time stimulation duration [14]. The combination of both vibrotactile and thermal stimuli is used in the generation of haptic sensation [15]. In spite of its capability of representing color identification and discrimination, thermal tactile displays alone are not well-suited to present rich information, because of their low spatial resolution and their lower transition-sensing time between on/off states [16].

*2.2. Electrotactile*

Electrotactile stimulation uses electrical current flow from electrodes embedded in the device to deliver stimuli to the sensing nerves of the skin, mimicking pressure or vibration without any actual mechanical actuators involved [17]. A 4 × 4 electrotactile matrix called SmartTouch is developed to selectively stimulate the Merkel's discs and Meissner corpuscles. The sensation is generated by electrodes that run electrical current pulses of 1–3 mA with a duration of 0.2 ms through the nerves of the skin [18]. A sensory substitution system that employs electrotactile and vibrotactile displays was developed [19]. There are also oral electrotactile displays. An array of 7 × 7 tactual actuators fabricated on polyimide-based flexible film was placed on the roof of the mouth, to deliver electrotactile stimulation with relatively low stimulation intensities [20]. Although electrotactile systems are structurally simple and easily controlled, challenges remain in spatial resolution, safety, comfort level, and power consumption.

*2.3. Mechanical*

Mechanical tactile stimulation is the most commonly used method to create a sensation, not only because the mechanoreceptors tend to respond to direct and physical mechanical stimuli easily, but also because of the finer spatial resolution as compared with thermal and electrotactile stimulation.

2.3.1. Static Indentation

Braille is the most well-known example of static tactile information acquisition. Braille can be printed on paper or produced on a refreshable Braille reader, in which the dots of each Braille cell are driven up and down by an array of stacked piezoelectric bending beam actuators. Braille is an excellent means of providing text information to those who are Braille-literate, as long as the information is available electronically as text, or can be scanned using optical character recognition. The most commonly-used technology for conveying tactile graphics is embossing of images on paper (e.g., thermoform or microcapsule paper) [21]. Refreshable Braille displays offer a quasi-static Braille display on a refreshed basis by employing extruded pins driven by piezoelectric bimorph actuators. This provides an opportunity for intensive information delivery using an electronic device that is more compact as compared with printed Braille books.

2.3.2. Vibration

Skin nerves tend to sense vibration easier than they sense static indentation [22]. Fast adapting Pacinian corpuscles are primarily responsible for vibrotactile perception in human skin. The lower thresholds of force and displacement for vibration, as compared with static indentation, make vibrotactile stimulation a widely and well-researched candidate for tactile displays. Thus, technologies have been developed to deliver vibrations for tactile display. As human beings, we can distinguish successive pulses with a time gap of 5 ms [23], which is even better than our vision system, in which the minimum time gap is 25 ms [24]. This means that vibration can be utilized to create rhythm or patterns for information with complex meaning. Further variation in vibro-rhythm can be realized by changing the amplitude and frequency.

2.3.3. Surface Acoustic Waves

Surface acoustic waves (SAW) generated by SAW transducers can stimulate the skin to provide a sensation of continuous roughness. Both passive and active sensation capture of SAW tactile devices have been reported [25–27]. Ultrasonic motors are used for generating vibration that can be directly sent to the user's finger, which is described as passive type SAW transducers. In contrast, active tactile transducers utilize standing waves of a SAW and friction shift to form vibration.

2.3.4. Electrorheological and Magnetorheological Fluids

Electrorheological (ER) and magnetorheological (MR) fluids are special classes of materials that can respond to the electrical field and magnetic field, respectively. They are both colloid suspensions with dielectric or ferromagnetic particles (1–100 μm) that are sensitive to electric or magnetic potential. Under normal conditions, when there is no electric or magnetic stimulus to the materials, the ER or MR fluids remain in liquid form. Upon application of the electric or magnetic field, the particles align themselves nearly parallel to the direction of the fields. This causes the viscosity to change, and the liquids become solid gels as the field applied increases. Such properties have been used to make tactile displays [28–30].

## 3. Vibrotactile Actuators for Tactile Feedback

Vibrotactile displays have been researched and developed during the past few decades, because mechanoreceptors sense vibration sensations more easily and rapidly than they sense quasistatic mechanical stimuli. The actuators used to generate vibration range from large-scale electric motors to MEMS-scale hydraulic pumps. Each has advantages and drawbacks for tactile displays.

### 3.1. Rotary Electromagnet Actuators

Rotary DC motors [31] are utilized to produce a vibrotactile sensation. The motors rotate when a DC current is applied. An off-centered mass affixed to the output shaft of the motor (often referred as an eccentric mass) offers the vibration. The feeling of vibration created by these motors varies linearly with the voltage or current applied. A small voltage creates a small and slow vibration, whereas a large applied voltage generates a strong and fast vibration. This type of actuator is the most commonly used vibrational actuator in toys, game controls, and virtual reality tactile devices (Axonvr). The benefits of rotary actuators include their cost-effectiveness, relatively strong vibration, and relatively lower requirements for electronics. Their drawbacks include their slow response time, which is usually in the range of tens of milliseconds. In addition, because the rotation it generates is non-directional, it is also not suitable for high-quality precision tactile feedback. Also, the size of the rotary motor is usually relatively large, which makes it a poor candidate for high-resolution tactile displays.

### 3.2. Linear Electromagnetic Actuators

Linear electromagnetic actuators (LEA) are another main way of using electromagnetic inputs to generate vibration. When a current passes through the conductive wires wrapped to form a coil, an electromagnetic field is generated. That field either pushes or drags a permanent magnet inside the coil, depending on its physical orientation and the direction in which the current flows into the coil; the motion of the magnet, in turn, causes a tactile vibration. LEA actuators are also commonly used in mobile phones, because of their low cost and the appropriateness of their compact size for the mobile phone scale. The actuator only works at the resonance frequency of the system. Although the LEA reacts faster than a rotary motor, it is still rather slow for a fast response tactile system. In addition, making the actuator can be complicated.

### 3.3. Electroactive Polymer

Electroactive polymers (EAP) are a group of polymers that change shape or size when an electric field is applied. When they change shape or size, a vibration is formed. Electroactive polymer actuators provide quasistatic millimeter-scale actuation, and are small enough to be arrayed with a pitch of a few millimeters, but they typically have actuation times on the order of tenths of seconds to tens of seconds [32,33]. The EAP is robust, but slow in refresh rate. In addition, a voltage higher than 300 V is normally needed to activate the actuator, which is not ideal in a portable tactile device.

### 3.4. Shape Memory Alloy

Shape memory alloy (SMA) actuators are metal alloys that remember their original shapes; their shapes change under the response to a temperature change, for example, via Joule heating. This effect happens because of the reversible phase changing inside the alloy. Though SMA tactile displays have been developed to provide large displacement and high force, their slow response times and large power consumptions make real-time vibrational graphical tactile display almost impossible [34–36].

*3.5. Carbon Nanotube*

As one of the promoted smart materials, carbon nanotube (CNT) plays an important role in providing actuation for tactile sensation. With the ultra-strong $sp^2$ carbon–carbon bond, carbon nanotubes (CNTs) have a great combination of high mechanical (5-fold stiffness vs steel, 10-fold strength vs carbon fiber) [37–40], electronic (current density is 1000 times higher than copper) [41], and thermal conducting properties [42]. This high aspect ratio, lightweight [43], and stable material can also be synthesized and fabricated into various sizes and shapes, as well as form composite networks with other functional materials [44,45]. All these advantages make CNTs an ideal candidate for making high performance mechanical, electrical, and even thermal tactile actuators [46–49].

Generally, the role CNTs play at actuation are utilizing the high electrical conductivity of CNTs to sense signals and transfer energy; realizing large stroke and fast response actuation by stretching, twisting, and bending the CNT based structures; or a combination of both. Baughman, et al. [50] have first demonstrated the electromechanical actuators based on sheets of single-walled CNTs, indicating CNT networks are a high potential durable media for various applications, such as low voltage-activated microcantilevers for medical catheters, or temperature insensitive material for jet engines. Lima, et al. [48] have demonstrated strong and fast hybrid CNT artificial muscles powered by electrically, chemically, and photonically activated guests.

A few approaches have been developed for tactile CNT actuators with different mechanisms. By adapting the high mechanical strength and high conductivity of MWCNT, Wang et al. [51] have developed a PDMS/MWCNT coplanar electrode layer to overcome crosstalk effects in the tactile array (Figure 2). Other approaches, such as activating CNT composites via different, have also shown promising results for tactile applications. Camargo et al. [52] have demonstrated a tactile device powered by optically activating the liquid crystal–carbon nanotube (LC–CNT) composite Braille dots (Figure 3). Pyo et al. [53] have demonstrated a CNT–PDMS composite tactile sensor, with the advantages of low-cost, batch production.

Due to a scaling degradation, large-scale CNT networks have much lower performance compared with ideal individual CNTs [46,47]; recent research has demonstrated that transfer of weak van der Waals CNT contacts into larger covalent junctions via appropriate electrical pulses is a promising method to overcome the challenge [54], which could further optimize the performance of CNT tactile actuators.

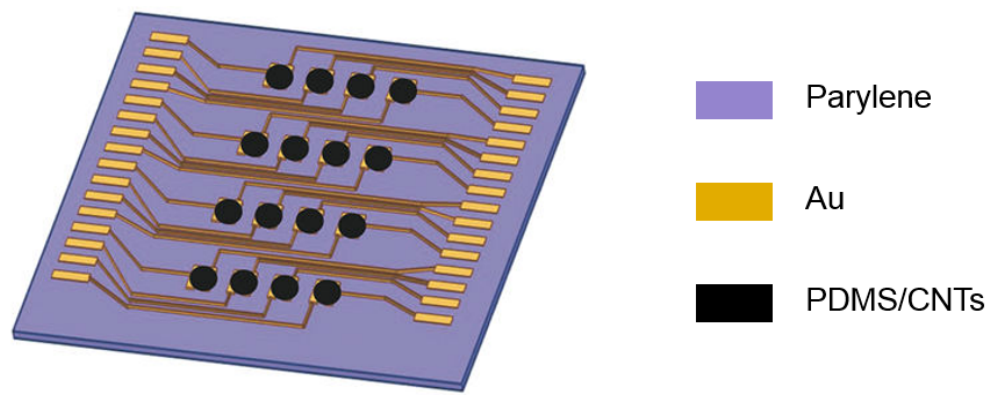

**Figure 2.** Assembly of PDMS/MWCNT-based tactile sensor array with coplanar electrodes. Adapted from [51].

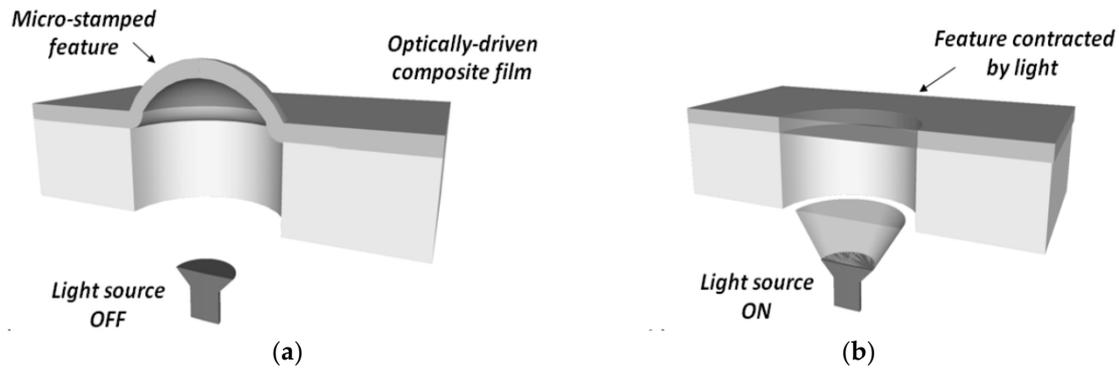

**Figure 3.** Schematic concept of carbon nanotube (CNT) composite film driven by the light source to form "blister" shape and flat shape for actuation. (**a**) Default state: the light source is OFF. The feature is a high-relief dot; (**b**) the feature flattens when it is light-actuated. This actuation is reversible when the light is turned off. Adapted from [52].

*3.6. Piezoelectric Actuators*

The piezoelectric effect was discovered in 1880 by two physicists [55]. It describes the phenomenon that in a certain group of materials, an electrical potential is generated when a mechanical load (pressing or squeezing) is applied on the material. In most crystals like metal, the unit cell that is the minimum repeating structure is symmetric. In contrast, in piezoelectric crystal structures, the unit cell is not symmetric. Piezoelectric crystals are electrically neutral in their initial state. They can have an electric polarization when no load is applied (being both ferroelectric and piezoelectric) or no electric polarization when no load is applied (being purely piezoelectric). When a mechanical load is applied, the positive and negative charges separate, generating an electrical potential across the piezoelectric material [56–59]. This process is also reversible. When a voltage is applied to the opposing faces of the piezoelectric material, the material needs to rebalance the electrical charges inside it, which causes a mechanical deformation. In order to create the piezoelectric effect in piezoceramics, in which the piezoelectric crystal grains are randomly oriented, the material needs to be heated to high-temperature under a strong electric field in a process called poling. The heat allows more free movement of the molecules, and the poling directions of each grain are forced into nearly the same direction under the strong electric field. After the poling process, the piezoelectric effect is obtained in the treated material.

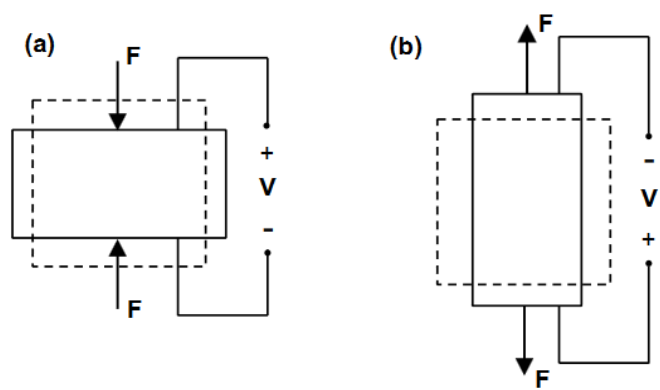

**Figure 4.** Schematic diagrams showing the piezoelectric effect. The dashed outline represents the object's original shape; the solid outline represents the object's deformed shape.

Frequently used piezoelectric materials include piezoelectric ceramics, such as lead zirconate titanate (PZT), and piezoelectric polymers, such as polyvinylidene fluoride (PVDF). The crystal structure of piezoelectric ceramics is close to a cube, in which, for example, eight metal ions with two negative charges take positions at the eight corners, six oxygen ions with two native charges take positions at the six faces, forming tetragons inside the cube, and a metal ion with four positive charges is located at the center of the cube and the tetragon. When the temperature of the material is above a certain value, called the Curie temperature (which depends on the material), each crystal has a symmetric structure with no electric dipole. When the temperature goes below the Curie temperature, the crystal cube turns dissymmetric, resulting in an electric dipole in a certain direction. If such dipoles randomly distribute inside a material, the material does not show net polarization. However, piezoelectric ceramics are ferroelectric below the Curie temperature, with permanent polarization and deformation after being poled by an electric field. In this way, such materials have the ability to generate an electric field in response to strain, or to undergo strain in response to external electric field. Piezoelectric polymers, such as PVDF, are also widely used as electrical generators. The structure of PVDF depends on the structure of the molecular chain, resulting in different polymer phases. Similarly, when poled, PVDF is a ferroelectric material that exhibits the piezoelectric effect.

The relationship between the mechanical deformation and the voltage applied to the piezoelectric materials is defined as a piezoelectric coefficient, which is mathematically defined as in Equation (1):

$$d = \frac{strain\ developed}{applied\ electric\ field} \quad (1)$$

Three axes, termed as 1, 2, and 3, are used to identify directions in piezoceramic material. The terms 1, 2, and 3 represent the axis of X, Y, and Z in the spatial 3D set of axes. The axis 3 is defined as the axis that is parallel to the polarization direction during the poling process. In practice, the piezoelectric coefficient is described as $d_{ij}$, in which the subscripts $i$ and $j$ represent the poling direction of the piezoelectric material, and the direction of the mechanical strain, respectively. A larger $d_{ij}$ value means that the material has a greater mechanical deformation under a same electric field in the specific direction defined by $i$ and $j$. The most commonly used $d_{ij}$ are $d_{33}$ and $d_{31}$. The term $d_{33}$ describes a positive strain in the direction of its length when an electric field is in the same direction. The term $d_{31}$ describes a negative strain in the transverse direction as the electric field.

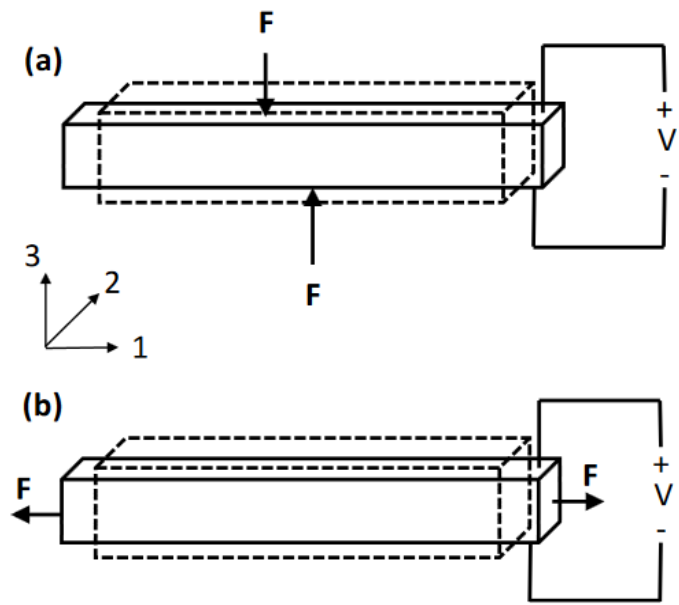

**Figure 5.** Schematic view of piezoelectric beams generating voltage in (**a**) 33 mode, and (**b**) 31 mode. The dashed outline represents the original shape; the solid outline represents the deformed shape.

The piezoelectric effect has been used to produce vibrotactile motion. Piezoelectric materials can be either solid ceramics or soft gel-like polymers, and they change shape when a voltage is applied. This process is reversible, so piezoelectric materials are also often used in sensors to detect mechanical deformation or the corresponding force applied on the material. Piezoelectric materials responds very quickly (microseconds) to electrical stimuli, and thus, are often used to produce high-frequency vibration.

3.6.1. Piezoelectric Bending Beam Actuators

The piezoelectric effect is found not only in piezoceramics, but also in polymers like PVDF (polyvinylidene fluoride). Piezoelectric polymer can provide a good pulling force, not a good pushing force, because of its mechanical flexibility. It also has a smaller piezoelectric coefficient compared to piezoelectric ceramic; thus, it is not suitable for applications in which high-frequency vibration with reasonable voltage is needed.

Piezoelectric actuators may be structured in various ways to produce extensional or bending actuation. When a single, long, thin plate or beam of a piezoceramic is poled in its thinnest dimension, it forms a piezoelectric unimorph actuator. When a voltage is subsequently applied across its thinnest dimension, the unimorph undergoes changes in length and thickness; in other words, it forms an extensional actuator. An alternative structure, a piezoelectric bimorph actuator, may be created by laminating two unimorphs together, so that the structure's thickness is doubled. The two unimorphs may be laminated, so that their polarizations are either parallel (called Y-poled) or antiparallel (called X-poled). When voltages are applied across the two layers of the bimorph, the bimorph can provide extensional actuation (like a unimorph actuator) or bending (when one layer is contracting, and the other is expanding under the influence of an applied voltage), depending on polarities of the applied voltages.

Piezoelectric bimorph bending-beam actuators may be used for vibrotactile applications. The vibration occurs when an alternating voltage is applied to the actuator. A device called the Optacon [60] was created in the 1970s to permit visually-impaired people to read text from a page without first translating the information into Braille. The system includes a camera that can be manually scanned across written text; an image of each character is then transferred into a pattern of vibrating piezoelectric beams in which the beam tips replicate the pattern of black and white of the character on the page. The piezoelectric beams are bimorph actuators that vibrate to deliver vibrotactile sensation. Although the Optacon is no longer in production, existing units are still in use by a group of people who find its capabilities indispensable. An alternative integrates piezoelectric bending beam actuators perpendicular to the tactile sensing plane, enabling large bending beam actuators to be tightly packed for fully 2D displays [61]. The large amount of piezoelectric material required potentially increases the cost of this type of architecture, as for other piezoelectric bending beam actuator systems.

3.6.2. Piezoelectric Extensional Actuators

Although most tactile displays that engage the piezoelectric effect utilize bending actuators, there are some tactile displays that use extensional piezoelectric actuators. Piezoelectric extensional actuators are not able to provide as large of a displacement as bending actuators, but with some motion amplification mechanisms, they can be compact and still efficient. Piezoelectric extension actuators and MEMS technology are also increasingly being leveraged to create tactile displays, as in [62–70]. Important challenges nonetheless remain, including spatial resolution, refresh rate, fabrication complexity, and cost.

*3.7. Pneumatic Actuation*

In pneumatic actuation, air pressure changes are used to provide direct or indirect vibration to the user's skin. The pressurized airflow directly causes a pressure or acts as a driver to move mechanical parts to interact with a finger [71,72]. Pneumatic actuator systems can be compact and light, but they require external air pumps to generate air pressure, and high-frequency output can be a problem for pneumatic tactile display.

## 4. Conclusions

Although refreshable 2D graphical interfaces have tremendous potential for conveying complex information, it is challenging to create actuators that are compact enough to be arrayed into an arbitrarily large number of rows and columns, while still being robust, easy to sense, and rapidly switchable. Each of the technologies presented in this chapter has its own advantages and disadvantages, certain types of actuator should be chosen in various applications in terms of the tradeoff between tactile feedback effectiveness, system complexity, power consumption, and cost, as shown in Table 2.

Research on high performance micro actuators is important, because they can be manufactured economically not only for assistive technology (e.g., interfaces through which people with blindness or low vision can perceive their environment), but also for enhancing the multi-sensory displays for private communication (e.g., silent, covert communication for military personnel), or advancing the tactile feedback in virtual reality, or providing motion and force in microrobotics. The main challenge in creating a small-sized actuator for high resolution tactile displays is that conventional macro actuators are large in size, making high resolution hard to implement, whereas MEMS actuators are small, making adequate force and displacement for human sensing difficult to attain. Suggested research work should investigate how actuators can bridge the gap between macroscale forces and microscale compactness to create effective tactile stimuli at the MEMS size scale.

Table 2. Comparison of common vibrotactile actuation types.

|  | Feedback Effectiveness | System Complexity | Size | Power Consumption | Response Speed | Cost |
|---|---|---|---|---|---|---|
| Rotary Motor | Bad | Good | Bad | Marginal | Good | Good |
| LEA | Marginal | Good | Bad | Marginal | Good | Good |
| EAP | Good | Marginal | Good | Bad | Bad | Marginal |
| SMA | Good | Bad | Good | Bad | Bad | Marginal |
| Piezoelectric | Good | Marginal | Marginal | Good | Good | Bad |
| Pneumatic | Good | Bad | Bad | Bad | Marginal | Bad |
| CNT | Good | Marginal | Bad | Marginal | Bad | Bad |

**Conflicts of Interest:** The authors declare no conflict of interest.

## References

1. Schultz, R.A.; Miller, D.C.; Kerr, C.S.; Micheli, L. Mechanoreceptors in human cruciate ligaments. A histological study. *J. Bone Jt. Surg. Am.* **1984**, *66*, 1072–1076.
2. Vallbo, Å.B.; Johansson, R.S. Properties of cutaneous mechanoreceptors in the human hand related to touch sensation. *Hum. Neurobiol.* **1984**, *3*, 3–14.
3. Blausen.com Staff. Medical Gallery of Blausen Medical 2014. *WikiJ. Med*. **2014**, doi:10.15347/wjm/2014.010, ISSN 20018762.
4. Choi, S.; Kuchenbecker, K.J. Vibrotactile display: Perception, technology, and applications. *Proc. IEEE* **2013**, *101*, 2093–2104.
5. Johnson, K.O. The roles and functions of cutaneous mechanoreceptors. *Curr. Opin. Neurobiol.* **2001**, *11*, 455–461.
6. Isseroff, R.G.; Sarne, Y.; Carmon, A.; Isseroff, A. Cortical potentials evoked by innocuous tactile and noxious thermal stimulation in the rat: Differences in localization and latency. *Behav. Neural Biol.* **1982**, *35*, 294–307.
7. Vallbo, Å.B.; Hagbarth, K.E. Activity from skin mechanoreceptors recorded percutaneously in awake human subjects. *Exp. Neurol.* **1968**, *21*, 270–289.
8. Lechelt, E.C. Tactile spatial anisotropy with static stimulation. *Bull. Psychon. Soc.* **1992**, *30*, 140–142.
9. Verrillo, R.T.; Fraioli, A.J.; Smith, X.R.L. Sensation magnitude of vibrotactile stimuli. *Atten. Percept. Psychophys.* **1969**, *6*, 366–372.
10. Verrillo, R.T. Vibrotactile thresholds for hairy skin. *J. Exp. Psychol.* **1966**, *72*, 47–50.
11. Ho, C.; Lukin, J.A. *Encyclopedia of Life Sciences*; Wiley-Blackwell: Hoboken, NJ, USA, 2000.

12. Downar, J.; Crawley, A.P.; Mikulis, D.J.; Davis, K.D. A multimodal cortical network for the detection of changes in the sensory environment. *Nat. Neurosci.* **2000**, *3*, 277–283.
13. Mao, N. Towards objective discrimination & evaluation of fabric tactile properties: Quantification of biaxial fabric deformations by using energy methods. In Proceedings of the 14th AUTEX World Textile Conference, Bursa, Turkey, 26–28 May 2014.
14. Yamamoto, A.; Cros, B.; Hashimoto, H.; Higuchi, T. Control of thermal tactile display based on prediction of contact temperature. In Proceedings of the 2004 IEEE International Conference on Robotics and Automation (ICRA'04), New Orleans, LA, USA, 26 April–1 May 2004; Volume 2, pp. 1536–1541.
15. Kron, A.; Schmidt, G. Multi-fingered tactile feedback from virtual and remote environments. In Proceedings of the IEEE 11th Symposium on in Haptic Interfaces for Virtual Environment and Teleoperator Systems (HAPTICS 2003), Los Angeles, CA, USA, 22–23 March 2003; pp. 16–23.
16. Jia, L.; Shi, Z.; Zang, X.; Müller, H.J. Watching a real moving object expands tactile duration: The role of task-irrelevant action context for subjective time. *Atten. Percept. Psychophys.* **2015**, *77*, 2768–2780.
17. Strong, R.M.; Troxel, D.E. An electrotactile display. *IEEE Trans. Man-Mach. Syst.* **1970**, *11*, 72–79.
18. Kajimoto, H.; Kawakami, N.; Tachi, S. Optimal design method for selective nerve stimulation and its application to electrocutaneous display. In Proceedings of the 10th Symposium on Haptic Interfaces for Virtual Environment and Teleoperator Systems (HAPTICS 2002), Orlando, FL, USA, 24–25 March 2002; pp. 303–310.
19. Kaczmarek, K.A.; Webster, J.G.; Bach-y-Rita, P.; Tompkins, W.J. Electrotactile and vibrotactile displays for sensory substitution systems. *IEEE Trans. Biomed. Eng.* **1991**, *38*, 1–16.
20. Imai, T.; Kamping, S.; Breitenstein, C.; Pantev, C.; Lütkenhöner, B.; Knecht, S. Learning of tactile frequency discrimination in humans. *Hum. Brain Mapp.* **2003**, *18*, 260–271.
21. Field, T.M. Touch therapy effects on development. *Int. J. Behav. Dev.* **1998**, *22*, 779–797.
22. Lee, M.H.; Nicholls, H.R. Review Article Tactile sensing for mechatronics—A state of the art survey. *Mechatronics* **1999**, *9*, 1–31.
23. Gescheider, G.A.; Wright, J.H.; Verrillo, R.T. *Information-Processing Channels in the Tactile Sensory System: A psychophysical and Physiological Analysis*; Psychology Press: Milton Park, UK, 2008.
24. Goldstein, H. Communication intervention for children with autism: A review of treatment efficacy. *J. Autism Dev. Disord.* **2002**, *32*, 373–396.
25. Kotani, K.; Ito, S.; Miura, T.; Horii, K. Evaluating tactile sensitivity adaptation by measuring the differential threshold of archers. *J. Physiol. Anthropol.* **2007**, *26*, 143–148.
26. Takasaki, M.; Kotani, H.; Mizuno, T.; Nara, T. Reproduction of tactile sensation using SAW tactile display. In Proceedings of the First IEEE Technical Exhibition Based Conference on Robotics and Automation (TExCRA'04), Tokyo, Japan, 18–19 November 2004; pp. 67–68.
27. Nara, T.; Takasaki, M.; Maeda, T.; Higuchi, T.; Ando, S.; Tachi, S. Surface Acoustic Wave (SAW) tactile display based on properties of mechanoreceptors. In Proceedings of the IEEE Virtual Reality, Yokohama, Japan, 13–17 March 2001; pp. 13–20.
28. Soto-Faraco, S.; Sinnett, S.; Alsius, A.; Kingstone, A. Spatial orienting of tactile attention induced by social cues. *Psychon. Bull. Rev.* **2005**, *12*, 1024–1031.
29. Soto-Faraco, S.; Ronald, A.; Spence, C. Tactile selective attention and body posture: Assessing the multisensory contributions of vision and proprioception. *Atten. Percept. Psychophys.* **2004**, *66*, 1077–1094.
30. Engel, J.; Chen, J.; Fan, Z.; Liu, C. Polymer micromachined multimodal tactile sensors. *Sens. Actuators A Phys.* **2005**, *117*, 50–61.
31. Boldea, I.; Nasar, S.A. Linear electric actuators and generators. *IEEE Trans. Energy Convers.* **1999**, *14*, 712–717.
32. Bar-Cohen, Y. Electroactive polymers for refreshable Braille displays. *SPIE Newsroom*, 2009, 11, doi:10.1117/2.1200909.1738.
33. Bicchi, A.; Scilingo, E.P.; Ricciardi, E.; Pietrini, P. Tactile flow explains haptic counterparts of common visual illusions. *Brain Res. Bull.* **2008**, *75*, 737–741.
34. Jairakrean, S.; Chanthasopeephan, T. Position control of SMA actuator for 3D tactile display. In Proceedings of the IEEE International Conference on Rehabilitation Robotics (ICORR 2009), Kyoto, Japan, 23–26 June 2009; pp. 234–239.
35. Matsunaga, T.; Makishi, W.; Totsu, K.; Esashi, M.; Haga, Y. 2-D and 3-D tactile pin display using SMA micro-coil actuator and magnetic latch. In Proceedings of the 13th International Conference on Solid-State

Sensors, Actuators and Microsystems, 2005. Digest of Technical Papers, Seoul, South Korea, 5–9 June 2005; Volume 1, pp. 325–328.
36. Mansour, N.A.; El-Bab, A.M.F.; Assal, S.F. A novel SMA-based micro tactile display device for elasticity range of human soft tissues: Design and simulation. In Proceedings of the 2015 IEEE International Conference on Advanced Intelligent Mechatronics (AIM), Busan, South Korea, 7–11 July 2015; pp. 447–452.
37. Robertson, D.H.; Brenner, D.W.; Mintmire, J.W. Energetics of nanoscale graphitic tubules. *Phys. Rev. B* **1992**, *45*, 12592–12595.
38. Yu, M.; Files, B.S.; Arepalli, S.; Ruoff, R.S. Tensile loading of ropes of single wall carbon nanotubes and their mechanical properties. *Phys. Rev. Lett.* **2000**, *84*, 5552–5555.
39. Xie, X.; Liu, S.; Yang, C.; Yang, Z.; Xu, J.; Zhai, X. The application of smart materials in tactile actuators for tactile information delivery. *arXiv* **2017**, arXiv:1708.07077.
40. Peng, B.; Locascio, M.; Zapol, P.; Li, S.; Mielke, S.L.; Schatz, G.C.; Espinosa, H.D. Measurements of near-ultimate strength for multiwalled carbon nanotubes and irradiation-induced crosslinking improvements. *Nat. Nanotechnol.* **2008**, *3*, 626–631.
41. Wei, B.Q.; Vajtai, R.; Ajayan, P.M. Reliability and current carrying capacity of carbon nanotubes. *Appl. Phys. Lett.* **2001**, *79*, 1172–1174.
42. Pop, E.; Mann, D.; Wang, Q.; Goodson, K.; Dai, H. Thermal conductance of an individual single-wall carbon nanotube above room temperature. *Nano Lett.* **2006**, *6*, 96–100.
43. Avouris, P. Molecular electronics with carbon nanotubes. *Acc. Chem. Res.* **2002**, *35*, 1026–1034.
44. Li, S.; Zhang, X.; Zhao, J.; Meng, F.; Xu, G.; Yong, Z.; Jia, J.; Zhang, Z.; Li, Q. Enhancement of carbon nanotube fibres using different solvents and polymers. *Compos. Sci. Technol.* **2012**, *72*, 1402–1407.
45. Beese, A.M.; Sarkar, S.; Nair, A.; Naraghi, M.; An, Z.; Moravsky, A.; Loutfy, R.O.; Buehler, M.J.; Nguyen, S.T.; Espinosa, H.D. Bio-inspired carbon nanotube—Polymer composite yarns with hydrogen bond-mediated lateral interactions. *ACS Nano* **2013**, *7*, 3434–3446.
46. Hill, F.A.; Havel, T.F.; Lashmore, D.; Schauer, M.; Livermore, C. Storing energy and powering small systems with mechanical springs made of carbon nanotube yarn. *Energy* **2014**, *76*, 318–325.
47. Liu, S.; Martin, C.; Lashmore, D.; Schauer, M.; Livermore, C. Carbon nanotube torsional springs for regenerative braking systems. *J. Micromech. Microeng.* **2015**, *25*, doi:10.1088/0960-1317/25/10/104005.
48. Lima, M.D.; Li, N.; de Andrade, M.J.; Fang, S.; Oh, J.; Spinks, G.M.; Kozlov, M.E.; Haines, C.S.; Suh, D.; Kim, S.J. Electrically, chemically, and photonically powered torsional and tensile actuation of hybrid carbon nanotube yarn muscles. *Science* **2012**, *338*, 928–932.
49. Chen, L.; Liu, C.; Liu, K.; Meng, C.; Hu, C.; Wang, J.; Fan, S. High-performance, low-voltage, and easy-operable bending actuator based on aligned carbon nanotube/polymer composites. *ACS Nano* **2011**, *5*, 1588–1593.
50. Baughman, R.H.; Cui, C.; Zakhidov, A.A.; Iqbal, Z.; Barisci, J.N.; Spinks, G.M.; Wallace, G.G.; Mazzoldi, A.; de Rossi, D.; Jaschinski, O.; et al. Carbon nanotube actuators. *Science* **1999**, *284*, 1340–1344.
51. Wang, L.; Peng, H.; Wang, X.; Chen, X.; Yang, C.; Yang, B.; Liu, J. PDMS/MWCNT-based tactile sensor array with coplanar electrodes for crosstalk suppression. *Microsyst. Nanoeng.* **2016**, *2*, doi:10.1038/micronano.2016.65.
52. Camargo, C.J.; Torras, N.; Campanella, H.; Marshall, J.E.; Zinoviev, K.; Campo, E.M.; Terentjev, E.M.; Esteve, J. Microstamped opto-mechanical actuator for tactile displays. In *Nano-Opto-Mechanical Systems (NOMS)*; International Society for Optics and Photonics: Bellingham, WA USA, 2011; Volume 8107, p. 810709.
53. Pyo, S.; Lee, J.I.; Kim, M.O.; Chung, T.; Oh, Y.; Lim, S.C.; Park, J.; Kim, J. Batch fabricated flexible tactile sensor based on carbon nanotube-polymer composites. In Proceedings of the Transducers & Eurosensors XXVII: The 17th International Conference on Solid-State Sensors, Actuators and Microsystems (TRANSDUCERS & EUROSENSORS XXVII), Barcelona, Spain, 16–20 June 2013.
54. Jung, H.Y.; Araujo, P.T.; Kim, Y.L.; Jung, S.M.; Jia, X.; Hong, S.; Ahn, C.W.; Kong, J.; Dresselhaus, M.S.; Kar, S.; et al. Sculpting carbon bonds for allotropic transformation through solid-state re-engineering of–sp 2 carbon. *Nature Commun.* **2014**, *5*, doi:10.1038/ncomms5941.
55. Liu, T. Passively-Switched Vibrational Energy Harvesters. Ph.D. Thesis, Northeastern University, Boston, MA, USA, 2017; ProQuest Dissertations Publishing: Ann Arbor, MI, USA, 2017; ProQuest Number10276237.
56. Liu, T.; Pierre, R.S.; Livermore, C. Passively-switched energy harvester for increased operational range. *Smart Mater. Struct.* **2014**, *23*, doi:10.1088/0964-1726/23/9/095045.

57. Liu, T.; Livermore, C. A compact architecture for passively-switched energy harvesters. *J. Phys. Conf. Ser.* **2015**, *660*, doi:10.1088/1742-6596/660/1/012090.
58. Liu, T.; Livermore, C. Passively tuning harvesting beam length to achieve very high harvesting bandwidth in rotating applications. In Proceedings of the PowerMEMS Atlanta, GA, USA, 2–5 December 2015; pp. 492–495.
59. Liu, T.; Liu, S, Xie, X.; Yang, C.; Yang, Z.; Zhai, X. Smart materials and structures for energy harvesters. *arXiv* **2017**, arXiv:1709.00493.
60. Velazquez, R.; Pissaloux, E.; Szewczyk, J.; Hafez, M. Miniature Shape Memory Alloy Actuator for Tactile Binary Information Display. In Proceedings of the 2005 IEEE International Conference on Robotics and Automation, Barcelona, Spain, 18–22 April 2005; doi:10.1109/robot.2005.1570302.
61. Shah, C.; Bouzit, M.; Youssef, M.; Vasquez, L. Evaluation of RU-Netra Tactile Feedback Navigation System for the Visually Impaired. In Proceedings of the 2006 International Workshop on Virtual Rehabilitation, New York, NY, USA, 29–30 August 2006; doi:10.1109/iwvr.2006.1707530.
62. Xie, X.; Zaitsev, Y.; Velásquez-García, L.F.; Teller, S.J.; Livermore, C. Scalable, MEMS-enabled, vibrational tactile actuators for high resolution tactile displays. *J. Micromech. Microeng.* **2014**, *24*, doi:10.1088/0960-1317/24/12/125014.
63. Yang, C.; Liu, S.; Xie, X.; Livermore, C. Compact, planar, translational piezoelectric bimorph actuator with Archimedes' spiral actuating tethers. *J. Micromech. Microeng.* **2016**, *26*, doi:10.1088/0960-1317/26/12/124005.
64. Xie, X.; Livermore, C. Passively self-aligned assembly of compact barrel hinges for high-performance, out-of-plane mems actuators. In Proceedings of the 2017 IEEE 30th International Conference on Micro Electro Mechanical Systems (MEMS), Las Vegas, NV, USA, 22–26 January 2017; pp. 813–816.
65. Xie, X.; Livermore, C. A pivot-hinged, multilayer SU-8 micro motion amplifier assembled by a self-aligned approach. In Proceedings of the IEEE 29th International Conference on Micro Electro Mechanical Systems (MEMS), Shanghai, China, 24–28 January 2016; pp. 75–78.
66. Xie, X.; Livermore, C. A high-force, out-of-plane actuator with a MEMS-enabled microscissor motion amplifier. *J. Phys. Conf. Ser.* **2015**, *660*, doi:10.1088/1742-6596/660/1/012026.
67. Yang, C.; Xie, X.; Liu, S.; Livermore, C. Resealable, ultra-low leak micro valve using liquid surface tension sealing for vacuum applications. In Proceedings of the 2017 19th International Conference on Solid-State Sensors, Actuators and Microsystems (TRANSDUCERS), Kaohsiung, Taiwan, 18–22 June 2017; doi:10.1109/transducers.2017.7994481.
68. Xie, X.; Zaitsev, Y.; Velasquez-Garcia, L.; Teller, S.; Livermore, C. Compact, scalable, high-resolution, MEMS-enabled tactile displays. In Proceedings of the Solid-State Sensors, Actuators, and Microsystems Workshop, Hilton Head Island, SC, USA, 8–12 June 2014; pp. 127–130.
69. Xu, J.; Xie, X.; Yang, C.; Shen, Z. Test and Analysis of Hydraulic Fracture Characteristics of Rock Single Crack. *Fluid Mech. Open Access* **2017**, *4*, doi:10.4172/2476-2296.1000164.
70. Xie, X. High Performance Micro Actuators for Tactile Displays. Ph.D. Thesis, Northeastern University, Boston, MA, USA, 2017; ProQuest Dissertations Publishing: Ann Arbor, MI, USA, 2017; ProQuest No. 10273384.
71. Yoo, J.; Yun, S.; Lim, S.; Park, J.; Yun, K.; Lee, H. Position controlled pneumatic tactile display for tangential stimulation of a finger pad. *Sens. Actuators A Phys.* **2015**, *229*, 15–22, doi:10.1016/j.sna.2015.03.023.
72. Kim, Y.; Oakley, I.; Ryu, J. Design and Psychophysical Evaluation of Pneumatic Tactile Display. In Proceedings of the 2006 SICE-ICASE International Joint Conference, Busan, South Korea, 18–21 October 2006; doi:10.1109/sice.2006.315347.